\documentclass[11pt,dvips]{article}
\usepackage{epsfig,times} 
\usepackage{picinpar}
\usepackage{wrapfig}
\usepackage{floatflt}
\makeatletter
\renewcommand{\section}{\@startsection{section}{1}{0in}
	{0.4\baselineskip}{0.1\baselineskip}{\Large\bf}}
\renewcommand{\subsection}{\@startsection{subsection}{2}{0in}
	{0.25\baselineskip}{-\baselineskip}{\large\bf}}
\renewcommand{\subsubsection}{\@startsection{subsubsection}{3}{0in}
	{0.1\baselineskip}{-\baselineskip}{\normalsize\bf}}

\makeatother
\begin{document}

%
\thispagestyle{myheadings}
%
\markright{OG 2.2.02}
\begin{center}
%
{\LARGE \bf A Search for Pulsed TeV Gamma Ray Emission from the Crab Pulsar}
\end{center}
\begin{center}
{\bf A.M.Burdett$^{1,2}$, 
I.H.Bond$^{2}$,
P.J. Boyle$^{3}$,
S.M.Bradbury$^{2}$, 
J.H.Buckley$^{4}$, 
D.A.Carter-Lewis$^{5}$, 
M.Catanese$^{5}$,
M.F.Cawley$^{6}$,
M.D'Vali$^{2}$,
D.J.Fegan$^{3}$, 
S.J.Fegan$^{1}$, 
J.P.Finley$^{7}$, 
J.A.Gaidos$^{7}$, 
G.H.Gillanders$^{8}$,
T.A.Hall$^{7}$, 
A.M.Hillas$^{2}$,
J.Kildea$^{3}$,
J.Knapp$^{2}$,
F.Krennrich$^{5}$, 
M.J.Lang$^{8}$,
S.LeBohec$^{5}$,
R.W.Lessard$^{1}$, 
C.Masterson$^{3}$, 
P.Moriarty$^{9}$,
J.Quinn$^{3}$, 
H.J.Rose$^{2}$,
F.W.Samuelson$^{5}$,
G.H.Sembroski$^{1}$, 
R.Srinivasan$^{7}$, 
V.V.Vassiliev$^{1}$,
T.C.Weekes$^{1}$}\\

{\it $^{1}$Whipple Observatory, Harvard-Smithsonian CfA, P.O. Box 97, Amado, 
AZ 85645-0097\\
$^{2}$Department of Physics, University of Leeds, Leeds, LS2 9JT,U.K.\\
$^{3}$Department of Exp. Physics, University College Dublin, Belfield, 
Dublin 4, Ireland\\
$^{4}$Department of Physics, Washington University, St. Louis,MO 63130\\
$^{5}$Department of Physics and Astronomy, Iowa State University, Ames, 
IA 50011\\
$^{6}$Department of Physics, National University of Ireland, Maynooth, 
Co. Kildare, Ireland\\
$^{7}$Department of Physics, Purdue University, West Lafayette, 
IN 47907\\
$^{8}$Department of Physics, National University of Ireland, Galway, Ireland\\
$^{9}$School of Science, Galway-Mayo Institute of Technology, Galway, Ireland}
\end{center}
\begin{center}
{\large \bf Abstract\\}
\end{center}
\vspace{-0.5ex}

%
%
We present the results of a search for pulsed TeV emission from the
Crab pulsar using the Whipple Observatory's 10~m gamma-ray telescope.
The direction of the Crab pulsar was observed for a total of
73.4~hours between 1994 November and 1997 March. Spectral analysis 
techniques were applied to search for the
presence of a gamma-ray signal from the Crab pulsar over the energy
band 250~GeV to 4~TeV. At these energies we do not see any evidence 
of the 33~ms pulsations present at lower energies from the Crab pulsar.  
The 99.9\% confidence level upper limit for pulsed emission above 
250~GeV is derived to be $4.8\times10^{-12}{\rm cm}^{-2}{\rm s}^{-1}$ or
$<$3\% of the steady flux from the Crab Nebula. These results imply a
sharp cut-off of the power-law spectrum seen by the EGRET instrument
on the {\em Compton Gamma-Ray Observatory}.  If the cut-off is
exponential, it must begin at 60~GeV or lower to accommodate these
upper limits.
%

\vspace{1ex}

\setcounter{footnote}{0}
\section{Introduction}
\label{section:intro}
The Crab pulsar/Nebula system is one of the most intensely studied
astrophysical sources with measurements throughout the electromagnetic
spectrum from the radio to the TeV energy band. In most regions of the
spectrum, the characteristic 33~ms pulsations of the pulsar are clearly 
visible. The pulse profile is unique amongst known pulsars in that it 
is aligned from radio to gamma-ray energies. The study of the pulsed 
emission in different energy ranges is of considerable importance to 
understanding the underlying emission mechanisms. In each of the two 
existing models which address the pulsed gamma-ray emission 
in detail, the outer gap model (Romani 1996) and the polar cap model
(Daugherty \& Harding 1982), the high energy flux arises from curvature
radiation of electron/positron pairs as they propagate along magnetic 
field lines in the
pulsar magnetosphere. The energy at which the pulsed flux begins to 
cut-off and the detailed spectral shape of the cut-off can help to 
distinguish between the two models. Given the detection of pulsations 
at energies up to 10~GeV by EGRET (Ramanamurthy et al. 1995) and the 
restrictive upper
limits above 300~GeV (e.g., Reynolds et al. 1993), the cut-off must 
occur in the $\sim$10 to 100~GeV energy range.  

\section{Analysis Technique and Selection Methods}
\label{figure:phase}
\begin{figwindow}[15,r,%
{\mbox{\epsfig{file=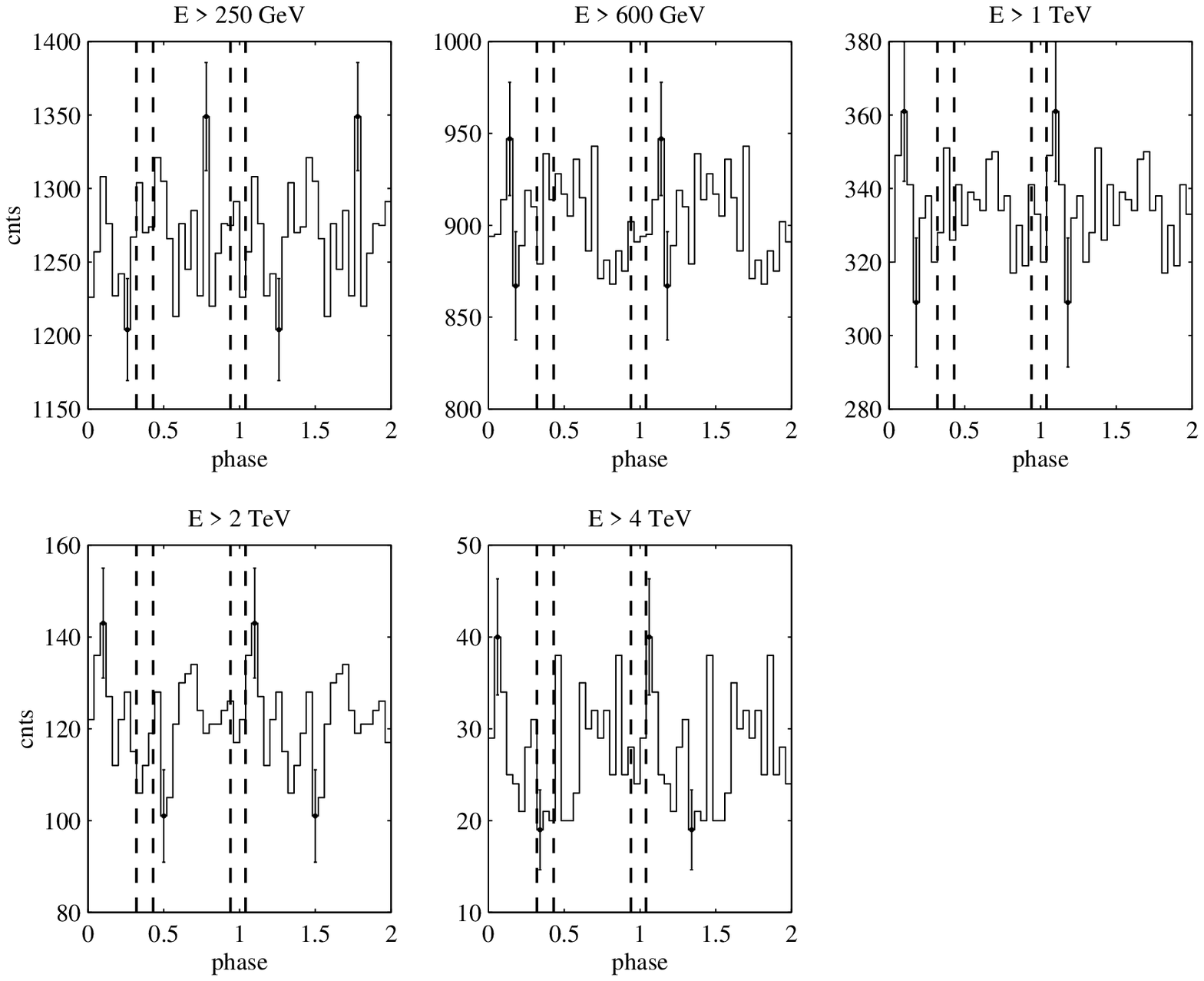,width=4.3in}}},%
{Search for TeV gamma rays from the Crab pulsar. The dashed
lines depict the EGRET main pulse and intrapulse phase ranges. Error
bars have been included on bins with the maximum and minimum number of
counts.}]
The standard gamma-ray selection method utilized by the Whipple
Collaboration is the Supercuts criteria (Reynolds et al. 1993). 
These criteria were optimized on contemporaneous
Crab Nebula data to give the best sensitivity to point sources. In
optimizing the overall sensitivity many showers below $\sim$400~GeV are
rejected.
In the context of a search for pulsed emission
from the Crab pulsar, which must have a low energy cut-off to
accommodate existing upper limits, it is clearlt desirable to include 
these low energy events. Accordingly, a modified set of cuts
(Moriarty et al. 1997), developed to
provide optimal sensitivity in the 200 to 400~GeV
region and referred to hereafter as Smallcuts, was used for the events
which failed the Supercuts pre-selection criteria. 
Simulations indicate that a combination of Supercuts and Smallcuts 
results in an energy threshold of $\sim$250~GeV. 
This threshold is the energy at which the
differential rate from a source with a spectral index equal to that of
the steady Crab Nebula reaches its peak. The effective collection
area is $\sim$2.7$\times$10$^8~{\rm cm}^2$.
Another selection process, known as Extended Supercuts (Mohanty et al. 
1998), was used in a search for pulsed emission over the energy band 
250~GeV to 4~TeV. This method is similar to Supercuts but instead scales 
the various cuts with the total shower {\em size} in ADC counts
of each event and retains 
approximately 95\% of gamma-ray events compared to approximately 50\% 
of gamma-ray events passed by the Supercuts criteria. 
The effective energy threshold for the analysis can be increased by 
applying a lower bound on the {\em size} of an image.
Lower bounds on the {\em sizes} of images of 500, 1000, 2000 and 5000 
digital counts lead to energy thresholds of 0.6, 1.0, 2.0 and 4.0~TeV,
respectively.
The arrival times of the Cherenkov events were registered by a GPS
clock with an absolute resolution of 250~$\mu$s. An oscillator,
calibrated by GPS second marks (relative resolution of 100~ns), was
used to interpolate to a resolution of 0.1~$\mu$s. The arrival times 
were transformed to the solar system barycentre and folded to
produce the phase with respect to the radio ephemeris obtained from
Jodrell Bank. An {\em optical} detection of the Crab pulsar was
successfully carried out using the 10m reflector with a photometer 
at its focus (Srinivasan et al. 1997), confirming the validity of the
timing of the experiment and the barycentering software used.
\end{figwindow}


\section{Results}
\label{section:discuss}
\label{figure:spectrum}
\begin{figwindow}[4,r,%
{\mbox{\epsfig{file=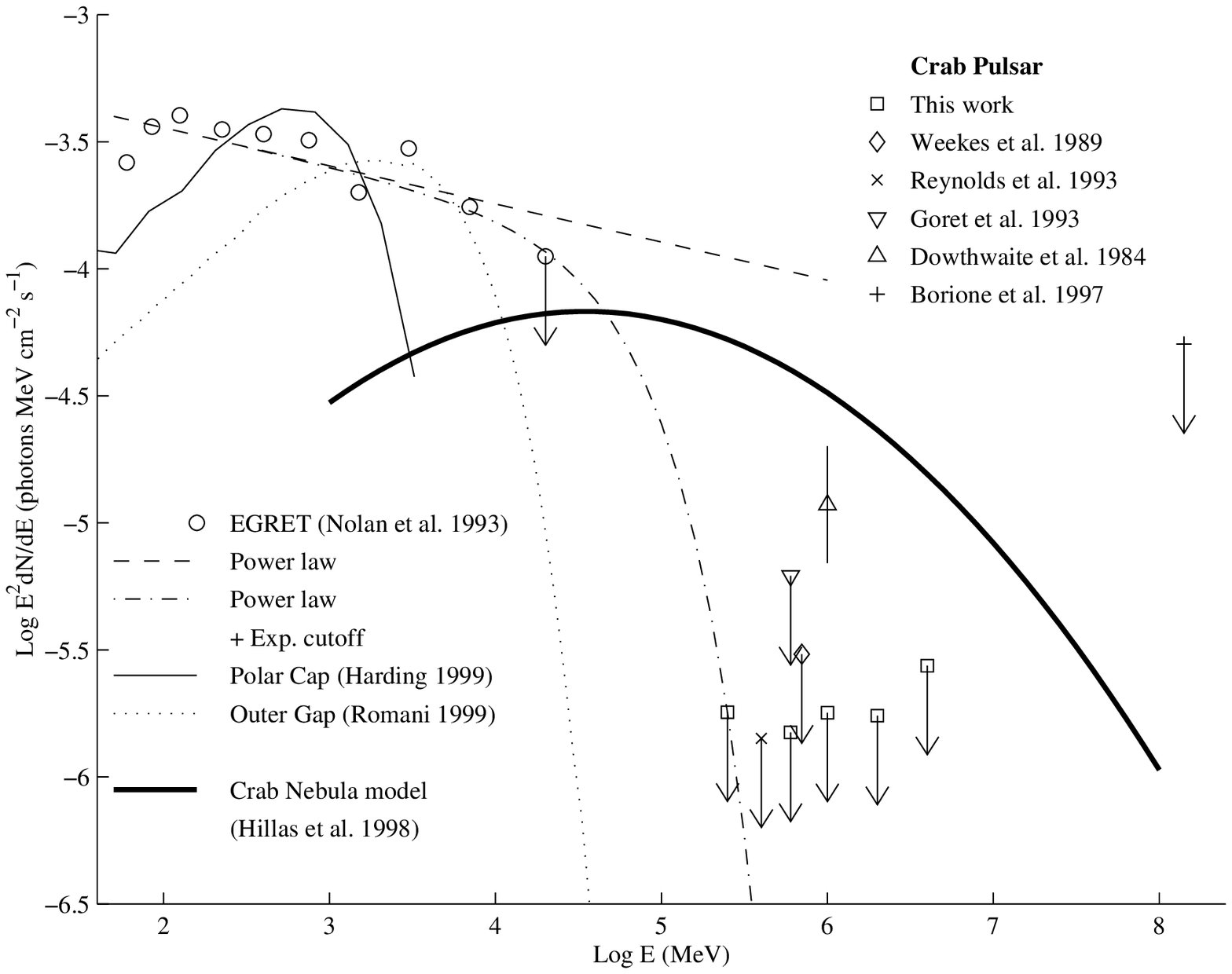,width=4.2in}}},%
{The pulsed photon spectrum of the Crab pulsar. The thin solid line 
is the polar cap model fit to the EGRET data (Harding 1999). The dotted
line is the outer gap model for the Vela pulsar (scaled to match the
EGRET Crab pulsar flux at peak intensity) and is included to indicate
the shape of the cut-off this model predicts (Romani 1999). The dashed 
line represents the power-law fit to the EGRET data (Nolan et al. 1993). 
The dot-dashed line represents Equation~1 with a cut-off energy $E_o = 60$
GeV. The thick solid curve depicts the model of unpulsed GeV - TeV emission 
from the Crab Nebula (Hillas et al. 1998)}]
The phases of the events passing cuts are shown in Figure 1. 
We find no evidence for emission pulsed at the radio period.
To calculate upper limits for pulsed emission we have used the pulse
profile seen at lower energies by EGRET (Fierro et al. 1998). That is, 
we assume emission occurs within the phase ranges of both the main 
pulse, phase 0.94-0.04, and the intrapulse, phase 0.32-0.43. The
number of events with phases within these intervals constitutes the
number of candidate pulsed events, $N_{on}$. $N_{off}$, an estimate of
the numbers of background events, is obtained by multiplying the
number of events with phases outside these pulse intervals by the
ratio of ranges spanned by the pulse and non-pulse regions. The
results are given in Table~1. The statistical
significance of the excess is calculated using the maximum likelihood
method of Li~\&~Ma (1983). The 99.9\% confidence level upper limits
calculated using the method of Helene (1983) are also given in
Table~1.
A few reports have described episodes of pulsed emission 
from the Crab pulsar at very high energies with time scales of several 
minutes (e.g., Bhat et al. 1986).
For this reason we have performed 
a run-by-run search for periodic emission from the Crab pulsar based on the 
above pulse profile. The results were consistent with the null hypothesis 
that there were no episodes of emission with timescales of the order of
28 minutes, the duration of each run.
\end{figwindow}

\section{Discussion}
Data taken with the Whipple Observatory's 10~m gamma-ray telescope
have been used to search for pulsations from the Crab pulsar above
250~GeV. We find no evidence for emission pulsed at the radio period
and upper limits on the integral flux have been calculated.

To model the pulsed gamma-ray spectrum, a function of the form
\begin{equation}
dN/dE = KE^{-\gamma}e^{-E/E_o}
\label{equation:powercutoff}
\end{equation}
was used, where $E$ is the photon energy, $\gamma$ is the photon
spectral index and $E_o$ is the cut-off energy. The source spectrum in
the EGRET energy range is well fitted by a power law with a photon
spectral index of $-2.15\pm0.04$ (Nolan et al. 1993). The pulsed upper
limit above 250~GeV reported here is $\sim$3 orders of magnitude below
the flux predicted by extrapolation of the EGRET power law.
Equation~1 was used to extrapolate the EGRET
spectrum to higher energies constrained by the TeV upper limit
reported here and indicates a cut-off energy $E_o \leq 60$ GeV for
pulsed emission (see Figure~2).


The sharpness and location of the spectral cut-off can be used to
discriminate between the emission models. Current observations and the 
derived cut-off given above indicate that the cut-off must lie in the 
10-60~GeV range. However, the upper limits reported here are well above
the flux predicted by the polar cap and outer gap models and offer no 
discrimination between them.

The outer gap model of Romani (1996) also predicts a component of emission
peaking around 1~TeV arising from the synchrotron-self-Compton mechanism. 
The non-detection of such a TeV peak does not rule out this mechanism 
since the flux produced is also dependent on the density of local soft
photons which are upscattered by the high energy electrons. 

\begin{table}
\caption{Selected events for periodic analysis. $N_{on}$ are the number
of events with phases within the EGRET pulse profile and $N_{off}$ are
the background estimated from events falling outside the EGRET pulse
profile.}
\label{table:table}
\vspace{.2cm}
\begin{tabular}{|l|ccccc|} \hline 
Selection   &$N_{on}$&$N_{off}$& Significance& Periodic Emission                     & Threshold\\
Method      &        &         & $\sigma$    & Upper Limit                           & (TeV)\\
	    &	     &         &             &(cm$^{-2}$s$^{-1}$)$\times$ 10$^{-13}$ &\\ \hline
Supercuts + Smallcuts                   & 6696     & 6636     & 0.65  & $<$48.2      & $\geq$ 0.25\\
Extended Supercuts ({\em size} $>$ 500) & 4709 	   & 4748     & -0.50 & $<$16.7      & $\geq$ 0.6\\
Extended Supercuts ({\em size} $>$ 1000)& 1738     & 1762     & -0.51 & $<$12.0      & $\geq$ 1.0\\
Extended Supercuts ({\em size} $>$ 2000)& 602      & 649      & -1.67 & $<$5.9       & $\geq$ 2.0\\
Extended Supercuts ({\em size} $>$ 5000)& 125      & 150      & -1.88 & $<$4.6       & $\geq$ 4.0\\
\hline
\end{tabular}
\end{table}

\begin{center}
{\large\bf Acknowledgements}
\end{center}
 We acknowledge the technical assistance of K. Harris,
T. Lappin, and E. Roache.  We thank A. Lyne and R. Pritchard for
providing the radio ephemeris of the Crab pulsar. This research is
supported by grants from the U.S.  Department of Energy, NASA,
Enterprise Ireland and by PPARC in the United Kingdom.

\vspace{1ex}
\begin{center}
{\Large\bf References}
\end{center}
Bhat, P.N., et al. 1986, Nature, 319, 127\\
Borione, A., et al. 1997, ApJ, 481, 313\\
Dowthwaite, J.C., et al. 1984, ApJ, 286, L35\\
Daugherty, J.K. \& Harding, A.K. 1982, ApJ, 252, 337\\
Fierro, J.M., et al. 1998, ApJ, 494, 734\\
Goret, P., et al. 1993, A\&A, 270, 401\\
Harding, A.K. 1999, private communication\\
Helene, O. 1983, Nucl. Instr. Meth, 212, 319\\
Hillas, A.M., et al. 1998, ApJ, 503, 744\\
Li, T.P. \& Ma, Y.Q. 1983, ApJ, 272, 317\\
Mohanty, G., et al. 1998, Astroparticle Physics, 9, 15\\
Moriarty, P., et al. 1997, Astroparticle Physics, 7, 315\\
Nolan, P.L., et al. 1993, ApJ, 409, 697\\
Ramanamurthy, P.V., et al. 1995, ApJ, 450, 791\\
Reynolds, P.T., et al. 1993, ApJ, 404, 206R\\
Romani, R.W. 1996, ApJ, 470, 469\\
Romani, R.W. 1999, private communication\\
Srinivasan, R., et al. 1997, in Towards a Major Atmospheric Cherenkov Detector - V (Kruger Park, South Africa), ed. O.C. de Jager, 51\\
Weekes, T.C., et al. 1989, ApJ, 342, 379\\

\end{document}